\documentstyle[prl,aps,twocolumn,epsf,rotate]{revtex}
\newcommand{\bff}[1]{{\mbox{\boldmath $#1$}}}
\begin{document}
\draft
\renewcommand\baselinestretch{1.32}

\title{Cranked Relativistic Hartree-Bogoliubov Theory:\\
Superdeformed Bands in the $A\sim 190$ Region}

\author{A. V. Afanasjev\cite{LAT}, J.\ K{\"o}nig and P.\ Ring}

\address{Physik-Department der Technischen Universit{\"a}t 
M{\"u}nchen, D-85747 Garching, Germany}

\date{\today}
\twocolumn[\hsize\textwidth\columnwidth\hsize\csname@twocolumnfalse\endcsname
\maketitle

\begin{abstract}
Cranked Relativistic Hartree-Bogoliubov (CRHB) theory is presented 
as an extension of Relativistic Mean Field theory with pairing 
correlations to the rotating frame. Pairing correlations are 
taken into account by a finite range two-body force of Gogny type 
and approximate particle number projection is performed by 
Lipkin-Nogami method. This theory is applied to the description 
of yrast superdeformed rotational bands observed in even-even
nuclei of the $A\sim 190$ mass region.  Using the well established 
parameter sets NL1 for the Lagrangian and D1S for the pairing
force one obtains a very successful description of data
such as kinematic ($J^{(1)}$) and dynamic ($J^{(2)}$) 
moments of inertia without any adjustment of new parameters.
Within the present experimental accuracy the calculated 
transition quadrupole moments $Q_t$ agree reasonably 
well with the observed data.
\end{abstract}

\vspace{0.3in}]%
\narrowtext

The investigation of superdeformation in different mass
regions still remains in the focus of low-energy nuclear
physics. Experimental data on superdeformed rotational
(SD) bands are now available in different parts of the
periodic table, namely, in the $A\sim 60$ \cite{Zn60}, 
80, 130, 150 and 190 \cite{SD-sys} mass regions. This 
richness of data provides the necessary input for a test 
of different theoretical 
models and the underlying effective interactions at 
superdeformation. Cranked relativistic mean field
(CRMF) theory developed in Refs.\ \cite{KR,KR.93,AKR.96}
represents one of such theories. It has been applied in 
a systematic way for the description of SD bands observed 
in the $A\sim 60$ and $A\sim 150$ mass regions. The pairing 
correlations in these bands are considerably quenched and at 
high rotational frequencies a very good description of 
experimental data is obtained in the unpaired formalism in 
most of the cases as shown in 
Refs.\ \cite{AKR.96,Hung,ALR.98,A60,Zn68}.

  On the contrary, pairing correlations have a considerable 
impact on the properties of SD bands observed in the $A\sim 190$ 
mass region and more generally on rotational bands at low spin. 
Different theoretical mean field methods have been applied for 
the study of SD bands in this mass region. These are the cranked 
Nilsson-Strutinsky approach based on a Woods-Saxon potential 
\cite{WCNWJ.91,WS.94}, self-consistent cranked Hartree-Fock-Bogoliubov
approaches based either on Skyrme \cite{Skyrme1,Skyrme2} 
or Gogny forces \cite{GognyFr,GognySp}. It was shown in different 
theoretical models \cite{WS.94,Skyrme1,Skyrme2,GognyFr,SWM.94} that 
in order to describe the experimental data on moments of inertia one 
should go beyond the mean field approximation and deal with 
fluctuations in the pairing correlations using particle number 
projection. This is typically done in an approximate way by the
Lipkin-Nogami method \cite{L.60,N.64,PNL.73}. With exception of 
approaches based on Gogny forces, special care should also be 
taken to the form of the pairing interaction. For example, 
quadrupole pairing has been used in addition to monopole 
pairing in the cranked Nilsson-Strutinsky approach \cite{WS.94}. 
A similar approach to pairing has also been used in projected shell 
model \cite{PSM}. Density dependent pairing has been used in 
connection to Skyrme forces \cite{Skyrme1}. These requires, however, 
the adjustment of additional parameters to the experimental data.

Cranked Relativistic Hartree-Bogoliubov (CRHB) theory presented 
in this article is an extension of cranked relativistic mean field 
(CRMF) theory to the description of pairing correlations in rotating
nuclei. A brief outline of this theory and its application 
to the study of several yrast SD bands observed in even-even nuclei 
of the $A\sim 190$ region with neutron numbers $N=110,112,114$ is 
presented below while more details (both of the theory and 
the calculations) will be given in a forthcoming publication. 

The theory describes the nucleus as a system of Dirac nucleons 
which interact in a relativistic covariant manner through the
exchange of virtual mesons \cite{SW.86}: the isoscalar scalar 
$\sigma$ meson, the isoscalar vector $\omega$ meson, and the 
isovector vector $\rho$ meson. The photon field $(A)$ accounts 
for the electromagnetic interaction. 

 The CRHB equations for the fermions in the rotating frame 
are given in one-dimensional cranking approximation by 
\begin{equation}
\pmatrix{ h - \Omega_x \hat{J}_x    & \hat{\Delta}  \cr
-\!\hat{\Delta}^* &     -h^* + \Omega_x \hat{J}^*_x \cr}
\pmatrix{ U_k \cr V_k } =
E_k \pmatrix{ U_k \cr V_k }
\label{CHF}
\end{equation}
where $h=h_D-\lambda$ is the single-nucleon Dirac Hamiltonian minus
the chemical potential $\lambda$ and $\hat{\Delta}$ is the pairing 
potential. $\hat{J_x}$ and $\Omega_x$ are the 
projection of total angular momentum on the rotation axis and 
the rotational frequency. $U_k$ and 
$V_k$ are quasiparticle Dirac spinors and
$E_k$ denote the quasiparticle energies. The variational 
principle leads to time-independent inhomogeneous Klein-Gordon 
equations for the mesonic fields in the rotating frame 
%
\begin{eqnarray}
\left\{-\Delta-({\sl\Omega}_x\hat{L}_x)^2 + m_\sigma^2\right\}~
\sigma(\bff r) & = & 
-g_\sigma \rho_s(\bff r) \nonumber \\
\,\,\,\,\,\,\,\,\,\,\,\,\,\,\,\,\,-g_2\sigma^2(\bff r)-g_3\sigma^3(\bff r)
\nonumber \\
\left\{-\Delta-({\sl\Omega}_x\hat{L}_x)^2+m_\omega^2\right\}
\omega_0(\bff r)&=&
g_\omega \rho_v^{is}(\bff r)
\nonumber \\
\left\{-\Delta-({\sl\Omega}_x(\hat{L}_x+\hat{S}_x))^2+
m_\omega^2\right\}~
\bff\omega(\bff r)&=&
g_\omega\bff j^{is}(\bff r)
\nonumber \\
\left\{-\Delta-({\sl\Omega}_x\hat{L}_x)^2+m_\rho^2\right\}
\rho_0(\bff r)&=&
g_\rho\rho_v^{iv}(\bff r)
\nonumber \\
\left\{-\Delta-({\sl\Omega}_x(\hat{L}_x+\hat{S}_x))^2+
m_\rho^2\right\}~
\bff\rho(\bff r)&=& g_\rho \bff j^{iv}(\bff r)
\nonumber \\
-\Delta~A_0(\bff r)&=&e\rho_v^p(\bff r)
\nonumber \\
-\Delta~\bff A(\bff r)&=&e\bff j^p(\bff r)
\label{KGeq}
\end{eqnarray}
where the source terms are sums of bilinear products of 
baryon amplitudes 
\begin{eqnarray}
\rho_s(\bff r) & = & \sum_{k>0} 
 (V_k^n(\bff r))^{\dagger} \hat{\beta} V_k^n (\bff r) 
+(V_k^p(\bff r))^{\dagger} \hat{\beta} V_k^p (\bff r) 
\nonumber \\
\rho_v^{is}(\bff r) & = & \sum_{k>0} 
 (V_k^n(\bff r))^{\dagger} V_k^n (\bff r) 
+(V_k^p(\bff r))^{\dagger} V_k^p (\bff r) 
\nonumber \\
\rho_v^{iv}(\bff r) & = & \sum_{k>0} 
(V_k^n(\bff r))^{\dagger} V_k^n (\bff r) 
-(V_k^p(\bff r))^{\dagger} V_k^p (\bff r) 
\nonumber \\
\bff j^{is}(\bff r) & = & \sum_{k>0} 
(V_k^n(\bff r))^{\dagger} \hat{\bff\alpha} V_k^n (\bff r) 
+(V_k^p(\bff r))^{\dagger} \hat{\bff\alpha} V_k^p (\bff r) 
\nonumber \\
\bff j^{iv}(\bff r) & = & \sum_{k>0} 
(V_k^n(\bff r))^{\dagger} \hat{\bff\alpha} V_k^n (\bff r) 
-(V_k^p(\bff r))^{\dagger} \hat{\bff\alpha} V_k^p (\bff r) 
\label{source}
\end{eqnarray}
The sums over $k>0$ run over all quasiparticle states corresponding
to positive energy single-particle states ({\it no-sea approximation}). 
In Eqs.\ (\ref{KGeq},\ref{source}), the indexes $n$ and $p$ indicate 
neutron and proton states, respectively, and the indexes $is$ 
and $iv$ are used for isoscalar and isovector quantities.
$\rho_v^p(\bff r)$, $\bff j^p(\bff r)$ in Eq.\ (\ref{KGeq})
correspond to $\rho_v^{is}(\bff r)$ and $\bff j^{is}(\bff r)$
defined in Eq.\ (\ref{source}), respectively, but with
the sums over neutron states neglected.  
 
The spatial components of the vector mesons give origin to 
a magnetic potential $\bff V (\bff r)$ which breaks  
time-reversal symmetry and removes the degeneracy between 
nucleonic states related via this symmetry \cite{KR.93,AKR.96}.
This effect is commonly referred as a {\it nuclear magnetism}
\cite{KR}. It is very important for a proper description of the 
moments of inertia \cite{KR.93}. Consequently, the spatial components 
of the vector mesons $\omega$ 
and $\rho$ are properly taken into account in a fully 
self-consistent way. Since the coupling constant of the 
electromagnetic interaction is small compared with the coupling 
constants of the meson fields, the Coriolis term for the Coulomb 
potential $A_0(\bff r)$ and the spatial components of the vector 
potential $\bff A(\bff r)$ are neglected.

\begin{figure}
\epsfxsize=\columnwidth
\centerline{\epsffile{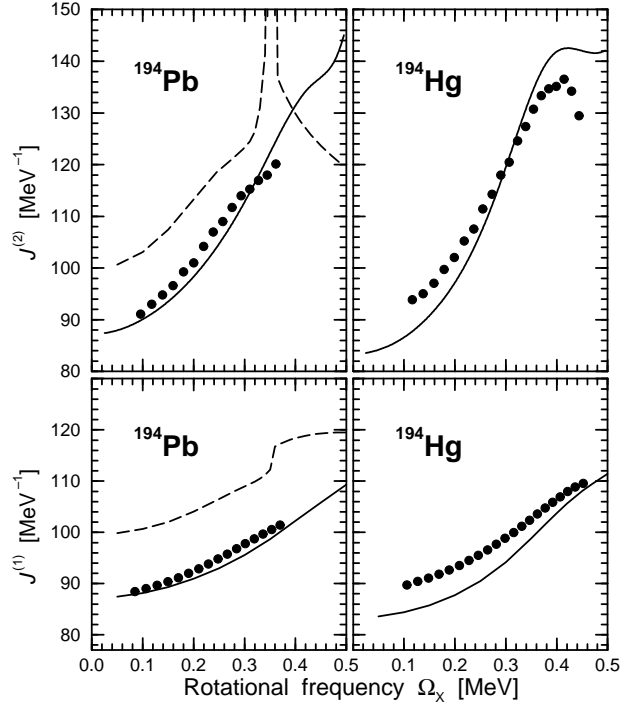}}
\vspace{-0.2cm}
\caption{Dynamic ($J^{(2)}$) (upper panels) and
kinematic ($J^{(1)}$) (bottom panels) moments of
inertia. The experimental values (full circles)
are compared with CRHB calculations (lines).
Solid and dashed lines show the results of
calculations with and without approximate
particle number projection performed by means
of Lipkin-Nogami method, respectively.
The experimental data are taken from
Refs.\ \protect\cite{Pb194a,Pb194b,Pb194c} ($^{194}$Pb)
and \protect\cite{Hg194a,Hg194b} ($^{194}$Hg).}
\label{pb94hg94}
\end{figure}

In the present version of CRHB theory, pairing correlations
are only considered between the baryons, because pairing is a
genuine non-relativistic effect, which plays a role only in
the vicinity of the Fermi surface. 
The phenomenological Gogny-type finite range interaction
\begin{eqnarray}
V^{pp}(1,2) &=& \sum_{i=1,2} e^{-[({\bff r}_1-{\bff r} _2)/\mu_i]^2} 
\label{eq1}\\
& & \times (W_i+B_i P^{\sigma}- H_i P^{\tau} - M_i P^{\sigma} P^{\tau})
\nonumber
\end{eqnarray}
with the parameters $\mu_i$, $W_i$, $B_i$, $H_i$ and $M_i$
$(i=1,2)$ is employed in the $pp$ (pairing) channel. The 
parameter set D1S \cite{BGG.91} has been used in 
the present calculations. This procedure requires no cutoff and 
provides a very reliable description of pairing properties in 
finite nuclei. In conjuction with relativistic mean field 
theory such an approach to the description of pairing correlations 
has been applied, for example, in the study of ground state 
properties \cite{LVR.98}, neutron halos \cite{PVLR.97}, and 
deformed proton emitters \cite{VLR.99}. In the present approach
we go beyond the mean field and perform an approximate particle number
projection before the variation by means of Lipkin-Nogami method 
\cite{L.60,N.64,PNL.73}. As illustrated in Fig.\ \ref{pb94hg94},
this feature is extremely important for a proper description of
the moments of inertia.

\begin{figure}
\epsfxsize=\columnwidth
\centerline{\epsffile{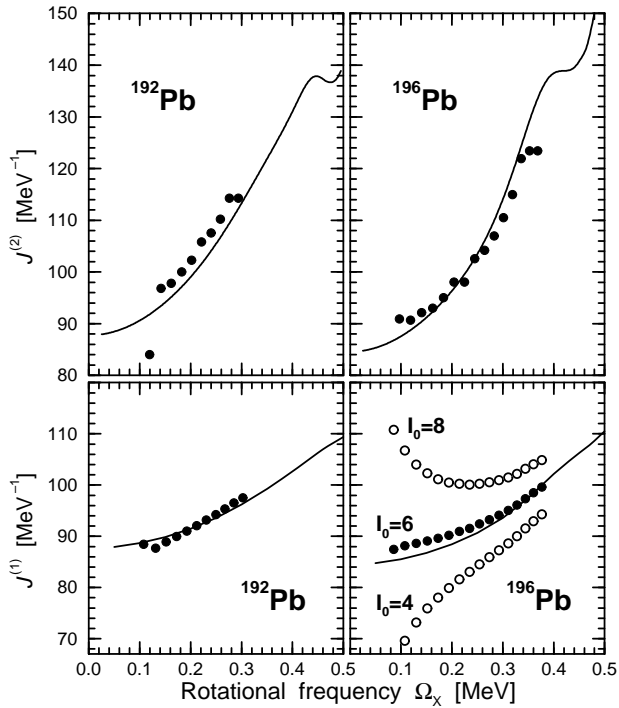}}
\vspace{-0.2cm}
\caption{The same as in Fig.\ \protect\ref{pb94hg94} but
for the yrast SD bands in $^{192,196}$Pb. The experimental 
data are taken from Refs.\ \protect\cite{Pb192a} 
($^{192}$Pb) and \protect\cite{Pb196a} ($^{196}$Pb).
The experimental $J^{(1)}$ values of the $^{192}$Pb band 
are shown for the spin value of $I_0=8$ for the 
lowest state in the SD band (the lowest transition with 
the energy 214.8 keV corresponds to a spin change 
$10 \rightarrow 8$). The experimental $J^{(1)}$ 
values of the $^{196}$Pb band are shown for three 
different values of $I_0$. The values being in 
best agreement with calculations are indicated by solid 
circles.}
\label{pb92pb96}
\end{figure}

   The present calculations have been performed with the NL1
parametrization \cite{NL1} of the relativistic mean field 
Lagrangian. The CRHB-equations are solved in the basis of 
an anisotropic three-dimensional harmonic oscillator in 
Cartesian coordinates. A basis deformation of $\beta_0=0.5$ 
has been used. All fermionic and bosonic states belonging 
to the shells up to $N_F=14$ and $N_B=16$ are taken into account 
in the diagonalisation and the matrix inversion, respectively. 
This truncation scheme provides reasonable numerical accuracy. 
For example, the increase of the fermionic basis up to $N_F=17$ 
changes the values of the kinematic moment of inertia $J^{(1)}$ and 
the transition quadrupole moment $Q_t$ by less than 1\%. 
The numerical errors for the total energy are even smaller. 

\begin{figure}
\epsfxsize=\columnwidth
\centerline{\epsffile{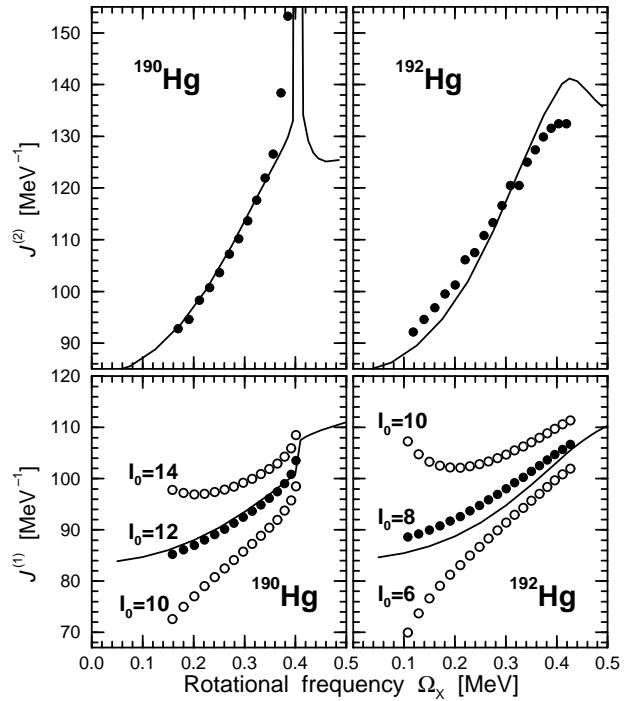}}
\vspace{-0.2cm}
\caption{The same as in Fig.\ \protect\ref{pb92pb96} 
but for the yrast SD bands in $^{190,192}$Hg. The experimental 
data are taken from Refs.\ \protect\cite{Hg190} ($^{190}$Hg) 
and \protect\cite{Hg192} ($^{192}$Hg).}
\label{hg90hg92}
\end{figure}

   Yrast SD bands in $^{194}$Pb and $^{194}$Hg are linked to the 
low-spin level scheme \cite{Pb194b,Pb194c,Hg194b}. In addition, 
there is a tentative linking of SD band in $^{192}$Pb \cite{Pb192a}.
These data provide an opportunity to compare with experiment in a 
direct way not only calculated dynamic ($J^{(2)}$) but also 
kinematic ($J^{(1)}$) moments of inertia. On the contrary, at present 
the yrast SD bands in $^{190,192}$Hg and $^{196}$Pb are not linked 
to the low-spin level scheme yet. Thus some spin values consistent 
with the signature of the calculated yrast SD configuration should 
be assumed for the experimental bands when a comparison is made with 
respect of the kinematic moment of inertia $J^{(1)}$.

  The results of such a comparison are shown in Figs.\ 
\ref{pb94hg94}, \ref{pb92pb96} and \ref{hg90hg92}. The theoretical 
$J^{(1)}$ values agree well with the experimental ones in the cases 
of linked SD bands in $^{194}$Pb and $^{194}$Hg and tentatively 
linked SD band in $^{192}$Pb. The comparison of theoretical and 
experimental $J^{(1)}$ values (see Figs.\  \ref{pb92pb96} 
and \ref{hg90hg92}) indicates that the lowest transitions
in the yrast SD bands of $^{190}$Hg, $^{192}$Hg and $^{196}$Pb
with energies 316.9, 214.4 and 171.5 keV, respectively, most 
likely correspond to the spin changes of $14^+ \rightarrow 12^+$, 
$10^+ \rightarrow 8^+$ and $8^+ \rightarrow 6^+$. If these spin
values are assumed, good agreement between theory and experiment 
is observed. Calculated and experimental values of the dynamic moment 
of inertia $J^{(2)}$ agree also  well, see Figs.\ \ref{pb94hg94}, 
\ref{pb92pb96} and \ref{hg90hg92}.

 The increase of kinematic and dynamic moments of inertia in this 
mass region can be understood in the framework of CRHB theory
as emerging predominantly from a combination of three effects: the 
gradual alignment of a pair of $j_{15/2}$ neutrons, the alignment of
a pair of $i_{13/2}$ protons at a somewhat higher frequency, and 
decreasing pairing correlations with increasing rotational 
frequency. The interplay of alignments of neutron and proton
pairs is more clearly seen in Pb isotopes where the calculated
$J^{(2)}$ values show either a small peak (for example, 
at $\Omega_x \sim 0.45$ MeV in $^{192}$Pb, see Fig.\ \ref{pb92pb96}) 
or a  plateau (at $\Omega_x \sim 0.4$ MeV in $^{196}$Pb, 
see Fig.\ \ref{pb92pb96}). With increasing rotational frequency, 
the $J^{(2)}$ values determined by the alignment in the neutron 
subsystem decrease but this process is compensated by the increase 
of $J^{(2)}$ due to the alignment of the $i_{13/2}$ proton pair. 
This leads to the increase of the total $J^{(2)}$-value at 
$\Omega_x \geq 0.45$ MeV. The shape of the peak (plateau) in $J^{(2)}$
is determined by a delicate balance between alignments in the proton 
and neutron subsystems which depends on deformation, rotational 
frequency and Fermi energy. 
For example, no increase in the total dynamic moment of inertia $J^{(2)}$ 
has been found in the calculations after the peak up to $\Omega_x=0.5$
MeV in $^{192}$Hg, see Fig.\ \ref{hg90hg92}.
It is also of interest to mention that the sharp increase in
$J^{(2)}$ of the yrast SD band in $^{190}$Hg is also reproduced 
in the present calculations. One should note that the 
calculations slightly overestimate the magnitude of 
$J^{(2)}$ at the highest observed frequencies. The possible 
reasons could be the deficiencies either of the Lipkin-Nogami 
method \cite{Mag93} or the cranking model in the band 
crossing region or both of them.

\begin{figure}
\epsfxsize=\columnwidth
\centerline{\epsffile{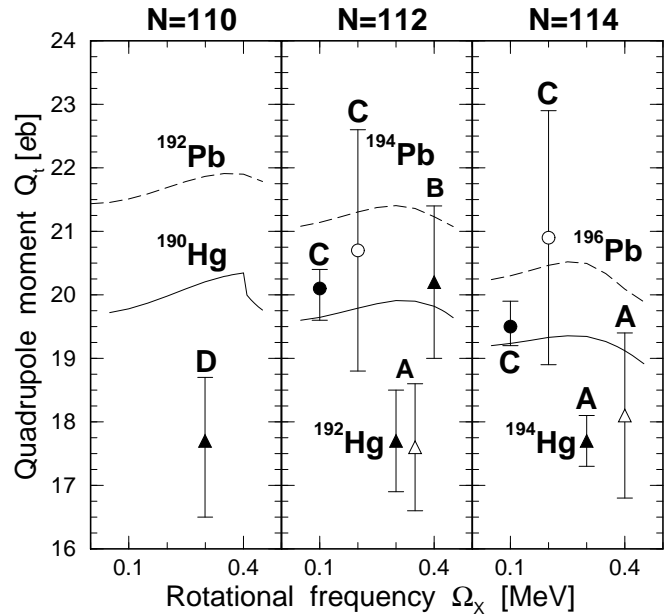}}
\caption{Measured and calculated transition quadrupole 
moments $Q_t$ of the yrast SD bands. Solid and dashed lines 
show the results of calculations for Hg and Pb isotopes, 
respectively. Experimental values are given by error
bars where the type of symbols corresponding to experimental
middle value of $Q_t$ indicates the isotope and type of
analysis used to extract the experimental value.
The following correspondence exists: circles - Pb isotopes,
triangles - Hg isotopes, open symbols - line shape analysis,
filled symbols - centroid shift. The experimental data are 
taken from Refs.\ \protect\cite{Hg9294-q} (exp. A),
\protect\cite{Hg92-q} (exp. B), \protect\cite{Pb9496-q} 
(exp. C), and \protect\cite{Hg90-q} (exp. D). Experimental 
data on $^{192}$Pb are not available.}
\label{qt}
\end{figure}

 The comparison between calculated and experimental absolute transition 
quadrupole moments $Q_t$ is less straightforward. This is because 
the uncertainties in absolute measured $Q_t$ values arising 
from the uncertainties in stopping powers can be as large as 15\% 
\cite{Hg9294-q}. Thus the comparison of $Q_t$'s values obtained in 
different experiments should be performed with some caution 
since systematic errors due to different stopping powers may be 
responsible for the observed differences. In addition, as illustrated 
in Fig.\ \ref{qt}, the experimental $Q_t$ values depend somewhat 
on the type of analysis (centroid shift or line shape) used when 
these quantities are extracted from the data.

 The results of CRHB calculations are compared with most recent 
experimental data in Fig.\ \ref{qt}. One can conclude that the 
results of calculations for absolute values of $Q_t$ will be 
within the `full' error bars if the 15\% uncertainty due to 
stopping powers would be taken into account (experimental data shown 
in Fig.\ \ref{qt} does not include these uncertainties). For the sake 
of simplicity we will not take into account these uncertainties 
in the subsequent discussion  and will concentrate mainly on the 
experimental data obtained with the same stopping powers. 
In Fig. \ref{qt} such data are indicated by the same capital
letters. While the calculated $Q_t$ values are close to the 
experimental values obtained with centroid shift and line shape 
analysis for Pb isotopes, most of experimental $Q_t$ values 
(with exception of exp. B) are overestimated in calculations in 
the case of Hg isotopes. One should note that the most recent 
experimental data on $^{192}$Hg is contradictory since two 
experiments (exp. A \cite{Hg9294-q} and exp. B \cite{Hg92-q}) 
give very different values of $Q_t$, see Fig.\ \ref{qt}. 
Definitely, the measurements of relative transition quadrupole 
moments between SD bands in Pb and Hg isotopes using the same stopping 
powers, which are not available nowadays, are needed to find out,
whether this discrepancy between calculations and experiment is due to
an inadequate theoretical description or the experimental problems 
quoted above. In the calculations, relative average quadrupole moments
$\Delta Q_t$ between yrast SD bands of Pb and Hg isotopes decrease
with increasing neutron number $N$ ($\Delta Q_t \approx 1.6$ $e$b, 
$\approx 1.4$ $e$b and $\approx 1.06$ $e$b for $N=110,112$ and 114, 
see Fig.\ \ref{qt}).

 Results of calculations indicate the general trend of decrease
of average $Q_t$ values with the increase of neutron number $N$ both 
for Pb and Hg isotopes. The results of the centroid shift analysis 
for $^{194,196}$Pb (exp. C) indicate a slight decrease in the 
$Q_t$ values with increasing $N$ consistent with theoretical 
results. Although the data on $^{192,194}$Hg (exp. A) indicate
similar values of $Q_t$ being in slight contradiction with 
theoretical results, definite conclusions are not possible at 
present due to the large error bars. In addition, with increasing 
rotational frequency $\Omega_x$ the calculated $Q_t$ values show 
initially a slight increase which is followed by a subsequent decrease. 
In the case of $^{190}$Hg this feature is hidden by the 
band crossing. The maximum $Q_t$ values within specific 
configurations are calculated at different frequencies $\Omega_x$ 
as a function of the neutron number $N$. With increasing $N$ 
the maximum $Q_t$ is reached at lower frequencies. Similar variations of 
$Q_t$ have been also observed in the cranked Hartree-Fock 
calculations with Skyrme forces \cite{Skyrme1}. Dedicated experiments 
aiming on the measurements of the variations of transition 
quadrupole moments $Q_t$ as a function of rotational frequency 
$\Omega_x$ are needed in order to confirm or reject these 
results.

  In conclusion, the Cranked Relativistic Hartree-Bogoliubov
theory has been developed and applied to the description of yrast 
SD bands observed in the $A\sim 190$ mass region. With an approximate 
particle number projection performed by the Lipkin-Nogami method, the 
rotational features of experimental bands such as kinematic and
dynamic moments of inertia are very well described in the 
calculations. Calculated values of transition quadrupole moments 
$Q_t$ are close to the measured ones, however, more accurate
and consistent experimental data on $Q_t$ is needed in 
order to make detailed comparisons between experiment and 
theory.

    A.V.A. acknowledges support from the Alexander von
Humboldt Foundation. This work is also supported in part
by the Bundesministerium f{\"u}r Bildung und Forschung
under the project 06 TM 875.



\end{document}